\documentclass[10pt,journal,compsoc]{IEEEtran}
\ifCLASSOPTIONcompsoc
  \usepackage[nocompress]{cite}
\else
  \usepackage{cite}
\fi
\ifCLASSINFOpdf
\else
\fi

\usepackage{booktabs}
\usepackage{amsmath}
\usepackage{graphicx}
\usepackage{subfigure}
\usepackage{amsfonts}

\begin{document}
\title{UPRec: User-Aware Pre-training for Recommender Systems}

\author{Chaojun~Xiao,
        Ruobing~Xie,
        Yuan~Yao,
        Zhiyuan~Liu,
        Maosong~Sun,
        Xu~Zhang,
        and~Leyu~Lin
        %Michael~Shell,~\IEEEmembership{Member,~IEEE,}
        %John~Doe,~\IEEEmembership{Fellow,~OSA,}
        %and~Jane~Doe,~\IEEEmembership{Life~Fellow,~IEEE}% <-this % stops a space
\IEEEcompsocitemizethanks{\IEEEcompsocthanksitem Chaojun Xiao, Yuan Yao, Zhiyuan Liu (corresponding author) and Maosong Sun are with the Department of Computer Science and Technology
Institute for Artificial Intelligence, Tsinghua University, Beijing, China, 100084. \protect \\
E-mail: xiaocj20@mails.tsinghua.edu.cn, yaoyuanthu@163.com,
\{lzy,sms\}@tsinghua.edu.cn
\IEEEcompsocthanksitem Ruobing Xie, Xu Zhang and Leyu Lin are with WeChat Search Application Department, Tencent, China, 100089. \protect \\
E-mail: xrbsnowing@163.com, \{xuonezhang,goshawklin\}@tencent.com
\IEEEcompsocthanksitem This work is finished during Chaojun Xiao's internship at WeChat Search Application Department, Tencent.
\IEEEcompsocthanksitem This work has been submitted to the IEEE for possible publication. Copyright may be transferred without notice, after which this version may no longer be accessible.
}% <-this % stops an unwanted space
}


\markboth{Journal of \LaTeX\ Class Files,~Vol.~14, No.~8, August~2015}%
{Shell \MakeLowercase{\textit{et al.}}: Bare Demo of IEEEtran.cls for Computer Society Journals}




\newcommand{\modelname}{UPRec}
\newcommand{\modelfullname}{\textbf{U}ser-aware \textbf{P}re-training for \textbf{Rec}ommendation}

\IEEEtitleabstractindextext{%
\begin{abstract}
%Sequential recommender systems aim to capture users' dynamic interests from their historical behaviour sequences, and have recently achieved significant progress with deep learning methods. 
Existing sequential recommendation methods rely on large amounts of training data and usually suffer from the data sparsity problem. To tackle this, %we propose a user-aware pre-trained method, called \modelfullname{}~(\modelname{}), which utilizes heterogeneous user information to construct self-supervised objectives in the pre-training stage. Specifically, we fuse the user attributes, the structured social graphs, and behaviour sequences via three simple and effective pre-training tasks. 
the pre-training mechanism has been widely adopted, which attempts to leverage large-scale data to perform self-supervised learning and transfer the pre-trained parameters to downstream tasks. However, previous pre-trained models for recommendation focus on leverage universal sequence patterns from user behaviour sequences and item information,
whereas ignore capturing personalized interests with the heterogeneous user information, which has been shown effective in contributing to personalized recommendation.
In this paper, we propose a method to enhance pre-trained models with heterogeneous user information, called \modelfullname{}~(\modelname{}). Specifically, \modelname{} leverages the user attributes and structured social graphs to construct self-supervised objectives in the pre-training stage and proposes two user-aware pre-training tasks. Comprehensive experimental results on several real-world large-scale recommendation datasets demonstrate that \modelname{} can effectively integrate user information into pre-trained models and thus provide more appropriate recommendations for users.
\end{abstract}

}

\maketitle

\IEEEdisplaynontitleabstractindextext

% \fi
\IEEEpeerreviewmaketitle

\IEEEraisesectionheading{\section{Introduction}\label{sec:introduction}}

\IEEEPARstart{W}{ith} the rapid development of various online platforms, large amounts of online items expose users to information overload. Recommender systems aim to accurately characterize users' interests and provide recommendations according to their profiles and historical behaviors. The application of recommendation systems makes it possible for users to obtain useful information efficiently, and thus has received great attention in recent years~\cite{zhang2019deep}. %Thus,  recommender systems have received great attention recently~\cite{}.

In many real-world scenarios, users' preference is intrinsically dynamic and evolving over time, which make it challenging to recommend appropriate items for users. 
%To address this issue, various methods have been proposed to make \emph{sequential recommendation} based on their chronological behaviours
\emph{Sequential recommendation} focuses on capturing users' long and short-term preferences and recommend the next items based on their chronological behaviours~\cite{hidasi2016session,kang2018self,li2017neural}. A main line of work attempts to obtain expressive user representations with sequential models, such as recurrent neural network~\cite{hidasi2016session,wu2017recurrent}, convolutional neural network~\cite{tang2018personalized}, and self-attention~\cite{kang2018self}. And some researchers seek to enhance the neural sequential models with rich contextual information, such as item attributes and knowledge graphs~\cite{hidasi2016parallel,huang2019taxonomy,zhang2019feature}.

These works achieve promising results in generating personalized recommendations. However, they rely on sufficient user behavior data for training and usually suffer from data sparsity problem~\cite{song2019autoint,yao2020self}. The similar problem also exists in the field of natural language processing (NLP). To tackle this, many efforts have been devoted to conducting self-supervised pre-training from large-scale unlabelled corpus~\cite{devlin2019bert,liu2019roberta}. It has been proven that pre-trained models can effectively capture complicated sequence patterns from large-scale raw data and transfer the knowledge to various downstream NLP tasks, especially in the few-shot setting~\cite{brown2020language,raffel2020exploring}.

Inspired by the success of pre-trained language models in NLP, many researchers propose to utilize pre-trained models, especially the BERT (Bidirectional Encoder Representations from Transformers)~\cite{devlin2019bert}, to derive user representations from their behaviour sequences in recommendation tasks~\cite{chen2019bert4sessrec,yuan2020parameter,zhou2020s3}. Similar to the masked language model, these works pre-train the model with a cloze-style task, which randomly masks some items in the behaviour sequences and requires the model to reconstruct the masked ones~\cite{sun2019bert4rec}.
%Sun et al.~\cite{sun2019bert4rec} introduce a cloze style objective to train BERT for recommendation. 
Furthermore, researchers seek to conduct more effective pre-training with various learning mechanisms~\cite{xie2020contrastive,zhou2020s3,yuan2020parameter}. And some works attempt to leverage item side information in the pre-training stage, including item attributes~\cite{zhou2020s3} and knowledge graphs~\cite{zeng2020knowledge}. These works have shown that pre-trained models can capture complex sequence patterns and generate expressive user representations even with sparse data.

% 然而，绝大多数推荐预训练的工作仅仅直接把MLM任务迁移至用户行为序列，并没有系统地考虑丰富的异质用户信息。个性化推荐和NLP不同，每个用户的行为序列是基于用户自身喜好的，有共性更有个性，而不是像NLP有一个general common LM。在推荐系统中，对用户信息的关注是必须的、不可或缺的。
However, these works mainly focus on adopt the cloze-style task on behaviour sequences, but ignore the abundant heterogeneous user information. Different from language understanding in NLP, which focuses on learning general language knowledge, recommender systems should not only leverage universal sequence patterns but also capture the personalized interests of each user. Therefore, it is necessary and indispensable to exploit user information for pre-trained recommender systems.
% However, few of these works exploit user information for pre-trained recommender systems.
Previous works have shown that the user information contains rich clues which can help models capture users' interest and further facilitate the recommender systems to alleviate the data sparsity problem~\cite{tang2013social,liu2018social}. For instance, users in the same age group tend to like similar songs for music recommendation~\cite{leblanc1996music}, and users who are friends tend to have similar behaviours for social recommendation~\cite{mcpherson2001birds}. 
%Hence, considering user information in pre-training can lead to better personalized behaviour modelling and accordingly benefit the downstream applications.

\begin{figure}
    \centering
    \includegraphics[width=0.95\linewidth]{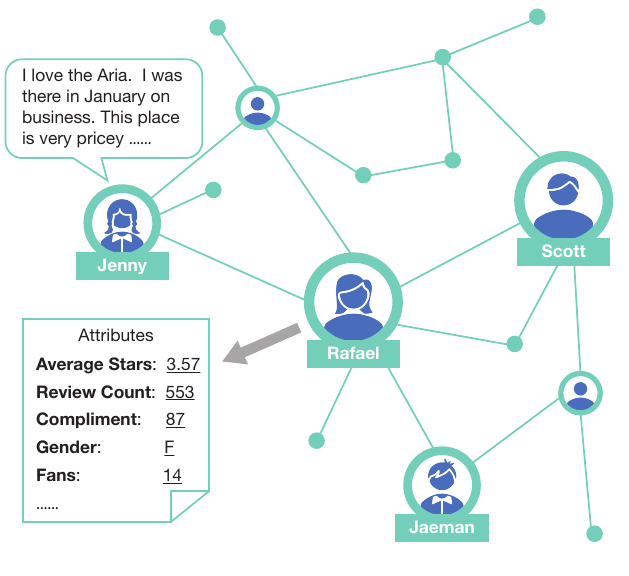}
    \caption{An example of heterogeneous user information from YELP. Each user can post their reviews for different items, and they are socially connected with each other to form a social graph. Besides, there are various attributes including profiles and behaviour properties for all users. The rich user information can help the recommender systems to better capture their preference.}
    \label{fig:example}
\end{figure}

Therefore, in this work, we propose to enhance pre-trained recommendation models with various user information. %, including social relations, user profiles, and user behaviour attributes. 
To this end, we need to tackle the problem of heterogeneous information integration. Figure~\ref{fig:example} shows an example from YELP online platform\footnote{https://www.yelp.com}. The user information is complicated and consists of various types of data, including sequential behaviour data, structured social graphs, and tabular user attributes. The formats of three types of data are quite different from each other, leading to three diverse semantic spaces. How to design special pre-training objectives to fuse the spaces together is an important problem.
%such a is an important problem.s structured social graphs armatio nnd numerical user profiles. How to design special objectives to fuse heterogeneous user information and behaviour sequences into the pre-trained model is an important problem.

To overcome this challenge, we propose \modelfullname{}~(\textbf{\modelname{}}), which use the same encoder to align the symbolic spaces of social graphs and user attributes with the semantic space of behaviour sequences under a pre-training framework. 
In particular, for the sequential behaviour data, we adopt the cloze-style Mask Item Prediction task to learn the items' and users' representations from bidirectional context following previous works~\cite{sun2019bert4rec,yuan2020parameter}.
Based on the representations, we propose two simple and effective user-aware pre-training tasks to leverage the two type of symbolic user information: (1)~User Attribute Prediction: We argue that the users' behaviour sequences can reflect various users' attributes to some extent. And in this task, we require the model to predict the users' attributes given the users' representations, which can help the model inject the attribute knowledge into the model. (2)~Social Relation Detection: The task is specially designed to incorporate social graphs into pre-training and aims to make the representations between socially connected users to be similar. Given representations of different users, we require the model to detect the social relations between them. 
% Besides, following previous works~\cite{sun2019bert4rec,yuan2020parameter}, we adopt the Mask Item Prediction task to learn the item representations from bidirectional context. Specifically, we randomly mask items in the behaviour sequences and require the model to predict the masked ones, which can help the model capture the complex sequence patterns.
Via these pre-training tasks, we can effectively fuse various kinds of user information and train a user-aware pre-trained model.

Moreover, to verify the effectiveness of \modelname{}, we conduct comprehensive experiments on two real-world sequential recommendation datasets from different domains. And we evaluate the performance of \modelname{} on the downstream tasks, user profile prediction. Experimental results show that both two user-aware pre-training tasks can help \modelname{} capture user interests more accurately, and achieve performance improvement. % Furthermore, ablation study and hyper-parameter sensitivity analysis are provided. 

To summarize, we make several noteworthy contributions in this paper: 
\begin{itemize}
    \item To the best of our knowledge, we are the first to systematically integrate heterogeneous user information, including user attributes, sequential behaviors and social graphs, under the pre-training recommendation framework.
    \item We propose two effective user-aware pre-training tasks: user attribute prediction and social relation detection, which possess plug and play characteristic and can be easily adopted in various recommendation scenarios.
    \item We perform comprehensive experiments two real-world recommendation datasets, and the experimental results demonstrate the effectiveness of our proposed model. The source code of this paper will be released to promote the improvements in recommender systems.
\end{itemize}

%The source code of this paper will be released to promote progress in recommendation once the paper is accpted.

\section{Related Work}
In this section, we will introduce previous works related to ours from the following three aspects, including general recommendation, sequential recommendation and pre-training for recommendation.

%\subsection{Social Recommendation}
%Utilizing social networks in recommendation has been studied for decades~\cite{ma2008sorec,tang2013social,yang2016social}. Homophily theory states that users' preference is influenced by their socially connected friends and behaviours between them tend to be similar~\cite{mcpherson2001birds,ibarra1993power}.

\subsection{General Recommendation}
Recommender systems aim to estimate user interests and recommend items that users may like~\cite{ricci2015recommender,zhang2019deep}. Existing recommendation models can be divided into two categories: collaborative filtering and content-based models. 

Collaborative filtering (CF) focuses on capturing user preference based on their historical feedback, such as clicks, likes. One typical class of CF is matrix factorization, which decomposes the user-item interaction matrix to obtain the user and item vectors, and the preference scores are estimated as the inner product between the user and item vectors~\cite{koren2015advances,koren2009matrix,guo2017deepfm}. Some works estimate the similarity between different items, and recommend items that are similar to ones the user has interacted with before~\cite{kabbur2013fism,linden2003amazon,sarwar2001item}. With the development of deep learning, various model architectures are introduced to learn the representations, such as multi-layer perceptions~\cite{he2017neural} and auto-encoder~\cite{sedhain2015autorec}.

Content-based models aim to integrate items' and users' auxiliary information into recommender models. These works mainly focus on enriching the items' or users' representation mainly by utilizing neural models to encode the side information, such as text~\cite{kim2016convolutional,wang2015collaborative}, images~\cite{kang2017visually,wang2017your}, and social graphs~\cite{liu2018social}.

\subsection{Sequential Recommendation}
Sequential recommendation aims to capture the users' dynamic preferences from their chronological user-item sequences~\cite{fang2020deep}.
Early works mainly rely on the Markov chain method, which predict the next item by estimating the item-item transition probability matrix~\cite{rendle2010factorization,quadrana2018sequence}. Further, some researchers employ the high order Markov chains to consider more items in the sequences~\cite{he2016fusing,he2017translation}.
%Early works are mainly based on effective machine learning algorithms such as Markov Chain and K-nearest neighbors~\cite{rendle2010factorization,quadrana2018sequence,he2016fusing}.

Recently, inspired by the powerful representation ability of various neural models, sequential neural models are widely adopted in the recommendation. For instance, some works propose to encode the user behaviour sequences with various recurrent neural networks, including Gated Recurrent Units (GRU)~\cite{hidasi2016parallel}, Long Short-Term Memory Network (LSTM)~\cite{wu2017recurrent} and other effective variants~\cite{quadrana2017personalizing,huang2018improving,ren2019repeatnet}.
%Inspired by the success of sequential neural models in natural language processing, many researchers introduce various neural networks for sequential recommendation, including recurrent neural networks~\cite{hidasi2016session,wu2017recurrent}, 
Besides, other powerful neural models are also introduced for recommendation. Tang et al.~\cite{tang2018personalized} utilize Convolutional Neural Networks to capture sequential patterns with both horizontal and vertical convolutional filters. Kang et al.~\cite{kang2018self} and Sun et al.~\cite{sun2019bert4rec} introduce the multi-head self-attention mechanism to model behaviour sequences. Though these approaches achieve remarkable results in sequential recommendation, they neglect the rich information about users. To tackle this issue, some works~\cite{rakesh2017probabilistic,liu2018social} incorporate social relations to the sequential recommendation. Despite the success of these models, the sufficient heterogeneous user information has not been fully utilized for user-item sequences modelling. 

\subsection{Pre-training for Recommendation}
Pre-training aims to learn useful representation from large-scale data, which will benefit specific downstream tasks. Recently, the pre-training mechanism achieves great success in many computer vision tasks~\cite{he2016deep,huang2017densely,simonyan2014very}, and natural language process tasks~\cite{devlin2019bert,liu2019roberta,yang2019xlnet}. In the field of recommender systems, pre-training has also received great attention. Early works attempt to apply pre-trained models to leverage side-information to enrich representations for users or items directly. According to the type of side-information, various pre-trained models are required. For instance, some researchers seek to utilize pre-trained word embeddings for textual data~\cite{zheng2017joint,gong2016hashtag}, pre-trained knowledge graph embeddings for knowledge graphs~\cite{zhang2016collaborative,huang2018improving,wang2018dkn} and pre-trained network embeddings for social graphs~\cite{chen2019n2vscdnnr,guo2018exploiting}. By leveraging the side-information, these approaches can construct expressive representations for users and items, thus achieve performance gain for recommender systems.

Recently, inspired by the rapid progress of pre-trained language models in natural language processing~\cite{devlin2019bert,liu2019roberta}, many efforts have been devoted to designing self-supervised pre-trained models to capture information from user behaviour sequences~\cite{zeng2020knowledge}. Sun et al.~\cite{sun2019bert4rec} and Chen et al.~\cite{chen2019bert4sessrec} propose to train the deep bidirectional encoder by predicting randomly masked item in sequences for sequential recommendation. Xie et al.~\cite{xie2020contrastive} further propose to utilize contrastive pre-training framework for sequential recommendation. Besides, some works attempt to utilize side-information of items, e.g., item attributes, in pre-training with mutual information maximization~\cite{zhou2020s3} and graph neural network~\cite{yang2020pre}. And Yuan et al.~\cite{yuan2020parameter} propose to fine-tune large-scale pre-trained network with parameter-efficient grafting neural networks. These works achieve significant improvement in user modeling and recommendation tasks. As these works mainly focus on utilizing item information or other recommendation tasks, they cannot be applied in this paper.

Different from previous works, we focus on constructing pre-training signals from user information, including user profiles and social relations. To the best of our knowledge, we are the first to enhance diverse user information in pre-training for the recommender system.

\section{Methodology}

\begin{figure*}
    \centering
    \includegraphics{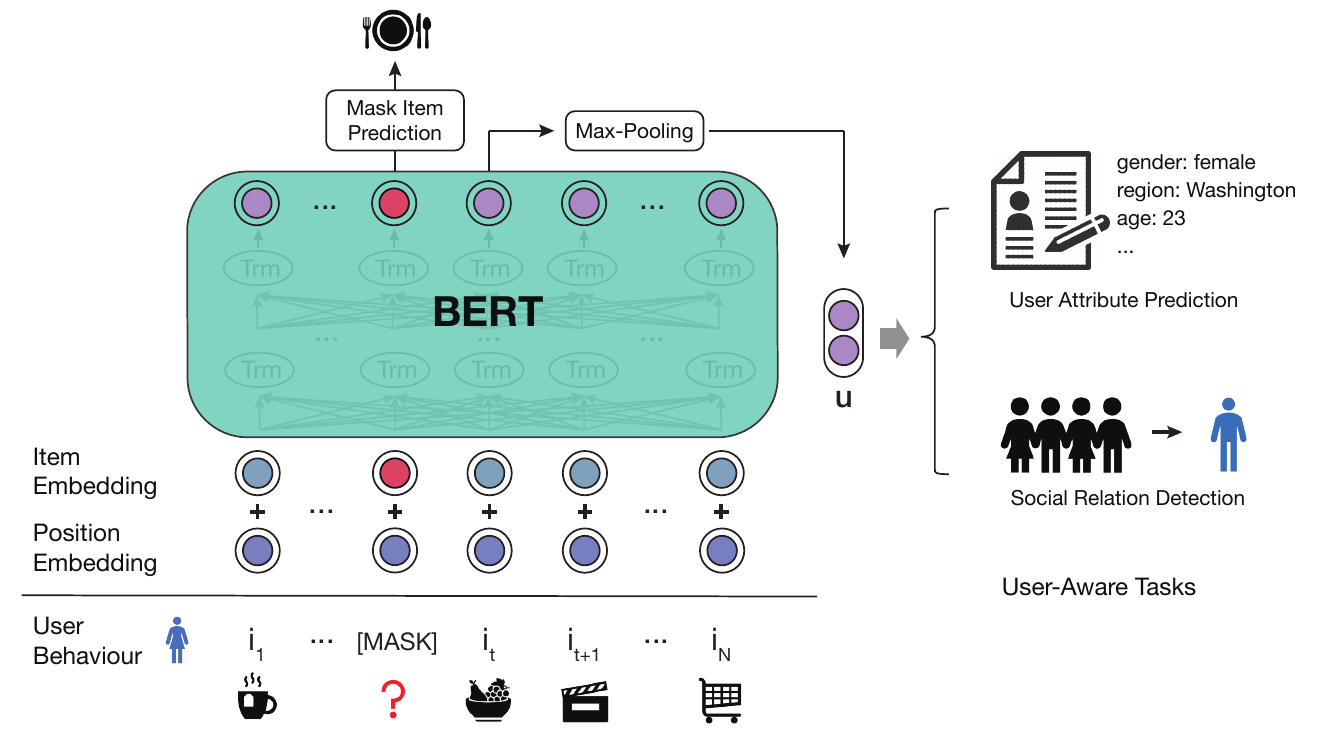}
    \caption{The architecture of \modelname{} in the pre-training stage. We adopt BERT as our encoder and utilize three tasks to pre-train \modelname{}: (1) Mask item prediction (MIP); (2) User attribute prediction (UAP); (3) Social Relation Detection (SRD). MIP allows the model to capture complex sequence patterns, and the two user-aware tasks can inject the user information into the representations.}
    \label{fig:my_model}
\end{figure*}

In this section, we will introduce our proposed user-aware pre-training framework for recommendation (\modelname{}), which incorporates various user information into the pre-trained model. The overview of \modelname{} is shown in Figure~\ref{fig:my_model}. We employ BERT~\cite{devlin2019bert} as our sequence encoder and utilize three objectives to pre-train the encoder. Following previous works~\cite{devlin2019bert,zhou2020s3,yuan2020parameter}, we adopt mask item prediction as our basic pre-training task to capture complex sequence patterns. Besides, in order to take full use of adequate user information, we propose two user-aware pre-training tasks: user attribute prediction and social relation detection, which leverage user attributes and social relations, respectively. %In the following sections, we will introduce these tasks in detail.

In the following sections, we will first introduce notations and our sequence encoder, BERT~\cite{devlin2019bert}. Then we will describe how we utilize three tasks to train \modelname{} in detail.

\subsection{Notations}
Let $\mathcal{U}$ denote the user set and $\mathcal{I}$ denote the item set. For each user $u \in \mathcal{U}$, we use $s_u = \{i^u_1, ..., i^u_{n}\}$ to represent his/her chronologically-ordered interaction sequence, where $i^u_j \in \mathcal{I}, 1 \leq j \leq n$, and $n$ is the sequence length. Let $\mathcal{R}_u$ denote the set of users who are socially connected with $u$. Besides, each user is associated with several attributes $\mathcal{A}_u = \{a^u_1,...,a^u_m\}$. The attributes can be very diverse. For instance, for the users from YELP platform, we can adopt the numerical average rating of all their posted reviews, and their gender, region as their attributes.

\subsection{Sequence Encoder: BERT}

\begin{figure}[t]
    \centering
    \includegraphics[width=0.6\linewidth]{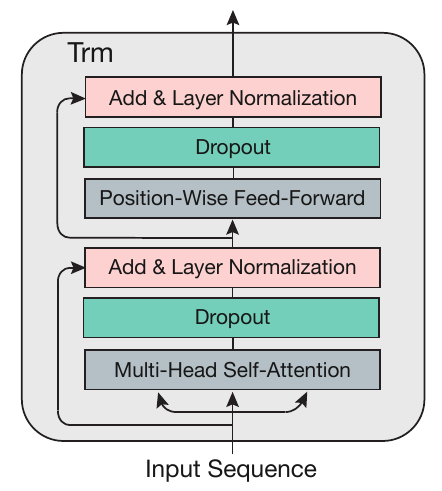}
    \caption{The architecture of the transformer layer, which consists of a multi-head self-attention layer and a point-wise feed-forward layer.}
    \label{fig:transformer}
\end{figure}

Sequential recommendation aims to exploit user chronological interaction sequence for next item recommendation. Inspired by the great success of pre-trained deep bidirectional transformers (BERT) in NLP~\cite{devlin2019bert,liu2019roberta}, many researchers begin to leverage BERT-based models to capture information from user behaviour sequences~\cite{sun2019bert4rec,chen2019bert4sessrec}. Following previous works, we adopt BERT as our basic module to encode the behaviour sequences. BERT is stacked by an embedding layer and $L$ bidirectional transformer layers. Figure~\ref{fig:transformer} presents the framework of the transformer layer. Each transformer layer consists of two sub-layer: multi-head self-attention layer and point-wise feed-forward layer. Then we will introduce the encoder in detail.

\subsubsection{Embedding Layer} 
In the embedding layer, the high dimensional one-hot representations of items are projected to low dimensional distributed representations with an item embedding matrix $\mathbf{M}$. Moreover, to make use of position information in sequence, learnable position embeddings are injected into the item representations. Formally, given the item sequence $\{i_1, ..., i_n\}$, we first map it into the embedding sequence $\{\mathbf{v}_1, ..., \mathbf{v}_n\}$, where $\mathbf{v}_i$ is $d$-dimensional vector. And the input representation is constructed by summing the item embeddings and position embeddings:
\begin{equation}
    \mathbf{h}_i^0 = \mathbf{v}_i + \mathbf{p}_i,
\end{equation}
where $\mathbf{p}_i \in \mathbb{R}^d$ is the position embedding for position index~$i$.

\subsubsection{Multi-Head Self-Attention Layer}
Compared with conventional neural models, self-attention mechanism is able to capture long distance dependencies from sequences. Thus, it has achieve promising results and is widely adopted in sequence modelling for both NLP and recommendation area. Moreover, multi-head mechanism allows the models to attend to information from multiple representation sub-spaces. Specifically, given the input hidden representation $\mathbf{H}^l$ from the $l$-th layer, the multi-head self-attention first project the input sequence into several vector sub-spaces, and then compute the output vectors with multiple attention heads:
\begin{align}
    \text{MultiHead}(\mathbf{H}^l) &= \text{Concat}(head_1, ..., head_h)\mathbf{W}^O, \\
    head_i &= \text{Attention}(\mathbf{H}^l\mathbf{W}_i^Q, \mathbf{H}^l\mathbf{W}_i^K, \mathbf{H}^l\mathbf{W}_i^V).
\end{align}
Here $h$ is the number of heads, $\mathbf{W}_i^Q$, $\mathbf{W}_i^K$, and $\mathbf{W}_i^V$ are trainable projection matrices for the $i$-th head, $\text{Concat}(\cdot)$ refers to concatenation operation and $\mathbf{W}^O$ is learnable parameters for output. And the attention function is implemented as scaled dot-product attention:
\begin{equation}
    \text{Attention}(\mathbf{Q}, \mathbf{K}, \mathbf{V}) = \text{softmax}(\frac{\mathbf{Q}\mathbf{K}^T}{\sqrt{d/h}})\mathbf{V},
\end{equation}
where query $\mathbf{Q}$, key $\mathbf{K}$, and value $\mathbf{V}$ are linear transformation from the same input hidden representation, $\sqrt{d/h}$ is the scaling factor. 

\subsubsection{Point-Wise Feed-Forward Layer} In addition to multi-head self-attention layer, each transformer layer also contains a fully connected feed-forward layer, which incorporate the model with non-linearity. In this layer, a feed-forward network is applied in each position separately and identically:
\begin{equation}
    \text{FFN}(\mathbf{h}_i^l) = \text{GELU}(\mathbf{h}_i^l\mathbf{W}^F_1+\mathbf{b}_1)\mathbf{W}^F_2+\mathbf{b}_2,
\end{equation}
where $\mathbf{W}^F_1$, $\mathbf{W}^F_2$, $\mathbf{b}_1$ and $\mathbf{b}_2$ are trainable parameters, and $\text{GELU}(\cdot)$ is the gaussian error linear unit activation function. The parameters are same for different positions in the same layer, but are different for different layers.

To avoid overfitting, a dropout operation is performed following each multi-head self-attention layer and point-wise feed-forward layer. Then a residual connection~\cite{he2016deep} and layer normalization operation~\cite{he2016deep} are employed to stabilize and accelerate the network training process.

\subsection{Pre-training Tasks}
Based on the above encoder, we further incorporate three pre-training tasks to enable the model to generate expressive sequence representations: Mask Item Prediction, User Attribute Prediction, and Social Relation Detection. The three objectives are optimized jointly.

\subsubsection{Mask Item Prediction}
Traditional sequential recommendation models usually utilize the left-to-right paradigm to train models to predict the next items. However, such unidirectional models will restrict the power of representations of items and sequences~\cite{sun2019bert4rec,yuan2020future}. Therefore, following previous works~\cite{sun2019bert4rec,zhou2020s3}, we adopt the mask item prediction (MIP) task in our method. MIP enables the model to generate item representation based on the context from both directions in the sequences and capture the complex sequence patterns. Specifically, when given a user-item interaction sequence $s=\{i_1,...,i_n\}$, we first randomly replace part of items with a special token \texttt{[MASK]}, and then the model is required to predict the masked ones based on their context. Formally, we first mask $p$ proportion items to get the inputs $\{i_1, ..., \texttt{[MASK]}, i_{t}, ..., i_n\}$, which are then fed into the BERT encoder to generate the hidden representations:
\begin{equation}
    \mathbf{H}^L = \text{BERT}(\{\texttt{[CLS]}, i_1, ..., \texttt{[MASK]}, i_{t}, ..., i_n, \texttt{[SEP]}\}).
    \label{eq:hidden}
\end{equation}
Here $\mathbf{H}^L$ is the hidden vectors of the sequences from the final layer, and \texttt{[CLS]} and \texttt{[SEP]} are special tokens used to mark the beginning and end of sequence, respectively. And the final hidden vectors corresponding to the \texttt{[MASK]} are fed into an output softmax function over the whole item set. And the loss is defined as the mean cross-entropy loss of each masked item:
\begin{equation}
    \mathcal{L}_{\text{MIP}} = -\frac{1}{|S_M|}\sum_{j \in S_M}-\text{log}P(i_{j}^p = i_{j}),
\end{equation}
where $S_M$ is the set of the positions of masked items, and $i_j^p$, $i_j$ are the predicted item and the original item at position $j$, respectively. Notably, in the fine-tuning stage, we adopt the task for sequential recommendation evaluation. In particular, we add the \texttt{[MASK]} to the end of the sequence, and require the model to recommend items based on the representation of the \texttt{[MASK]} token.

\subsubsection{User Attribute Prediction}
User attributes can provide sufficient fine-grained information about the users' preferences. And it is crucial to take full advantage of user attributes for recommendation. For instance, music tastes change over time, and different generations prefer different music~\cite{smith1994generational}. Therefore, we aim to inject the useful attributes information into the user representations. We argue that the users' behaviours can reflect the information about the users' attributes.
Specifically, we propose the user attribute prediction (UAP) task, which requires the model to predict the user attributes based on their interaction sequences.

Formally, given the final hidden representations $\mathbf{H}^L$ as in Equation~\ref{eq:hidden}, we first employ max-pooling operation to obtain the user representation:
\begin{equation}
    \mathbf{u} = \text{MaxPooling}(\mathbf{H}^L).
    \label{eq:user}
\end{equation}
Here $\mathbf{u}$ is the user representation. For different types of attributes, we employ different loss functions.
For numerical attributes, such as age and the average rating of all reviews, we formalize the task as a regression problem. We project the user representation to the predicted value with a linear layer and  minimize the Huber loss~\cite{huber1992robust}:
\begin{equation}
    \mathcal{L}_{\text{r}} = \frac{1}{|\mathcal{U}|} \sum_{u \in \mathcal{U}} z_u ,
\end{equation}
where $z_u$ is given by:
\begin{equation}
    z_u=\left\{\begin{array}{ll}
        0.5\left(a^p-a^u\right)^{2}  & \text { if }\left|a^p-a^u\right|< 1 \\
        \left|a^p-a^u\right|-0.5  & \text { otherwise }
\end{array}\right.,
\end{equation}
where $a^p$ and $a^u$ are the predicted value and ground truth value of the attribute. For discrete attributes, such as gender and region, we formalize the task as a classification problem.
Similar to the MIP task, we employ an output softmax function over the value set of the attribute, and define the loss as the mean cross-entropy loss:
\begin{equation}
    \mathcal{L}_\text{c} = \frac{1}{|\mathcal{U}|} \sum_{u \in \mathcal{U}} -\text{log}P(a^p = a^u).
\end{equation}

The overall loss of the UAP task is computed as the sum of all loss of different attributes:
\begin{equation}
    \mathcal{L}_\text{UAP} = \sum_{a \in \mathcal{A}_n} \mathcal{L}_\text{r} + \sum_{a \in \mathcal{A}_d} \mathcal{L}_\text{c},
\end{equation}
where $\mathcal{A}_n$ is the set of numerical attributes and $\mathcal{A}_d$ is the set of discrete attributes.

\subsubsection{Social Relation Detection}
Previous works demonstrate that users who are socially connected are more likely to share similar preferences~\cite{mcpherson2001birds}. And incorporating social relations in recommender systems can improve the performance of the personalized recommendation~\cite{ma2008sorec,fan2019graph}. Therefore, we formalize the task as a metric learning problem, and the goal is to create a vector space such that the distance of representations between friends is smaller than irrelevant ones. Formally, given the training data $\{\mathbf{u}_q, \mathbf{u}_c^+, \mathbf{u}_{c,1}^-, ..., \mathbf{u}_{c,m}^-\}$, where $\mathbf{u}_q$ is the query user, $\mathbf{u}_c^+ \in \mathcal{R}_u$ is the friend of the query user and $\mathbf{u}^-_{c,i}$ is the negative samples. We define the similarity between the query $\mathbf{u}_q$ and the candidate $\mathbf{u}_c$ as:
\begin{equation}
    \text{sim}(\mathbf{u}_q, \mathbf{u}_{c}) = -\left[\mathbf{w}_s^T\left(\mathbf{u}_q - \mathbf{u}_{c} \right)^2 + b_s \right],
\end{equation}
where the square notation indicates squaring each dimension of the vector. The similarity function can be regarded as a weighted L2 similarity with trainable $\mathbf{w}_s$ and $b_s$. And we optimize the loss function as the cross-entropy loss:
\begin{equation}
    \mathcal{L}_{\text{SRD}} = -\text{log}\frac{e^{\text{sim}(\mathbf{u}_q, \mathbf{u}_c^+)}}{e^{\text{sim}(\mathbf{u}_q, \mathbf{u}_c^+)} + \sum_{j=1}^m e^{\text{sim}(\mathbf{u}_q, \mathbf{u}_{c,j}^-)}}.
    \label{eq:loss_srd}
\end{equation}

For this task, the positive candidate users are provided explicitly, while the negative candidates need to be sampled from the whole user set. And how to select the negative samples is important for training a high-quality sequence encoder. Inspired by the previous works, we employ the in-batch negative strategy in this task. That is, we reuse the positive candidate from the same batch as negatives, which can make computation efficient and achieve great performance. Formally, we have $B$ user pairs $\{(\mathbf{u}_{1,q}, \mathbf{u}_{1,c}^+),...,(\mathbf{u}_{B,q}, \mathbf{u}_{B,c}^+)\}$ in a mini-batch. For each query user $\mathbf{u}_{j,q}$, $\mathbf{u}_{j,c}^+$ is his/her positive candidate and $\mathbf{u}_{i,c}^+ (i \neq j)$ are his/her negative candidates. Moreover, we argue that if two users are two-hop friends or have similar profiles, they are likely to become friends in the future. Therefore, to avoid introducing noise, we mask these negative candidate users, who are two-hop friends or have similar profiles with the query users.

\subsection{Training}
The process of \modelname{} consists of two steps. We first pre-train the encoder by optimizing the weighted sum of three tasks:
\begin{equation}
    \mathcal{L} = \lambda_1 \mathcal{L}_\text{MIP} + \lambda_2 \mathcal{L}_\text{UAP} + \lambda_3 \mathcal{L}_\text{SRD},
\end{equation}
where $\lambda_1$, $\lambda_2$ and $\lambda_3$ are hyper-parameters. In the fine-tuning stage, we employ the pre-trained parameters to initialize the encoder's parameters for downstream tasks. For the sequential recommendation task, we fine-tune the model by masking the last item of each sequence and adopt the negative log-likelihood of the masked targets to optimize the model. For the user profile prediction task, we use the hidden vector of the beginning token \texttt{[CLS]} to represent users, and then adopt the regression objective for numerical attributes and classification objective for discrete attributes.

\section{Experiment}
To verify the effectiveness of \modelname{}, we conduct experiments on two large-scale real-world datasets. Besides, ablation study and hyper-parameter sensitive analysis are provided to study whether \modelname{} works well in detail.
In order to evaluate the generalization ability of \modelname{}, we perform experiments on the user profile prediction tasks. The comprehensive analysis proves that \modelname{} can capture useful information from behaviour sequences and improve the performance of recommendation.

\subsection{Experimental Settings}

%\textbf{Datasets.}
\subsubsection{Datasets}

To evaluate our proposed model, we conduct experiments on two datasets collected from real-world platforms.
\begin{itemize}
    \item[(1)] \textbf{YELP}\footnote{https://www.yelp.com/dataset}: this is a large-scale dataset for business recommendation, which is collected from an online social network. Users make friends with each other and post reviews and ratings for items on the website. Following previous works~\cite{zhou2020s3}, we only use the interaction records after January 1st, 2019. And we treat the user metadata as attributes, including the number of compliments received by the users and the average rating of their posted reviews. 
    \item[(2)] \textbf{WeChat}\footnote{https://weixin.qq.com}: we build a new large-scale dataset from the largest Chinese social app, WeChat. We randomly select some users and collect their click behaviours in two weeks to build this dataset. Besides, we collect their gender, age, and regions as their attributes for the user attribution prediction task. The large-scale WeChat dataset contains tens of millions of interaction records and hundreds of thousands of social relations.
\end{itemize}

For these datasets, we group interaction records by users and sort them by timestamp to construct behaviour sequences.
Following previous works~\cite{zhang2019feature,rendle2010factorizing,zhou2020s3}, we only keep the $5$-core data, and filter out users and items with less than $5$ interaction records in the data preprocessing stage. We keep $90\%$ data for pre-training and sequential recommendation evaluation. The rest $10\%$ data are used for user profile prediction and social relation detection evaluation. The statistics of these datasets are shown in Table~\ref{tab:statistics_dataset}.

\begin{table}[h]
\small
\centering
\caption{The statistics of the datasets. The \#Users, \#Items, \#Rels, \#Interactions refer to the number of users, items, social relations and user-item interactions in each dataset.}
\begin{tabular}{lrrrr}
\toprule
Dataset & \multicolumn{1}{c}{\# Users} & \multicolumn{1}{c}{\# Items} & \multicolumn{1}{c}{\# Rels} & \multicolumn{1}{c}{\# Interactions} \\
\midrule
YELP    &  30,431  & 20,033   & 221,844 &  316,354                \\
WeChat  & 646,233  & 141,939  & 474,179 &  65,678,562             \\ 
\bottomrule
\end{tabular}
\label{tab:statistics_dataset}
\end{table}

\begin{table*}[ht]
%\small
\centering
\caption{Performance on the sequential recommendation task of different methods on two different real-world datasets. The best performance are denoted in bold. Note that, ``--" means the model does not converge.}
\resizebox{\textwidth}{!}{
\begin{tabular}{l|cccccc|cccccc}
\toprule
Dataset  & \multicolumn{6}{c|}{YELP}                     & \multicolumn{6}{c}{WeChat}                   \\ \midrule
Models   & HR@1 & HR@5 & HR@10 & NDCG@5 & NDCG@10 & MRR 
         & HR@1 & HR@5 & HR@10 & NDCG@5 & NDCG@10 & MRR \\ \midrule
GRU4Rec  & 09.86 & 37.66 & 57.22 & 23.81 & 30.13 & 24.02 
         & --    & --    & --    & --    & --    & --   \\ %(YELP 70.pkl)
Caser    & 14.57 & 40.54 & 56.97 & 27.81 & 33.12 & 27.75 
         & 11.64 & 37.64 & 57.10 & 24.67 & 30.95 & 25.11 \\
SASRec   & 14.17 & 38.98 & 53.71 & 26.85 & 31.62 & 26.73 
         & 17.32 & 48.89 & 66.05 & 33.50 & 39.05 & 32.35 \\ %(YELP 185.pkl)
BERT4Rec & 14.60 & 45.98 & 64.81 & 30.58 & 36.66 & 29.79 
         & 28.61 & 58.02 & 72.46 & 43.85 & 48.53 & 42.45 \\ \midrule % (source code)
$\text{\modelname{}}_{\text{w/o All}}$  
         & 15.04 & 46.31 & 65.50 & 30.93 & 37.14 & 30.25 
         & 29.57 & 64.62 & 79.36 & 47.91 & 52.70 & 45.47  \\
\modelname{}     & \textbf{16.96} & \textbf{49.04} & \textbf{68.81} & \textbf{33.24} & \textbf{39.63} & \textbf{32.31} 
          & \textbf{29.94} & \textbf{64.86} & \textbf{79.32} & \textbf{48.22} & \textbf{52.92} & \textbf{45.75} \\
% \midrule
% & \multicolumn{12}{c}{Cold Setting} \\ \midrule
% Caser    & 0.1065 & 0.2824 & 0.4138 & 0.1958 & 0.2381 & 0.2050 &      &      &       &        &         &     \\
% GRU4Rec  & 0.1998 & 0.5134 & 0.6855 & 0.3615 & 0.4172 & 0.3507 &      &      &       &        &         &     \\ %(YELP 24.pkl)
% SASRec   & 0.2321 & 0.5408 & 0.6870 & 0.3928 & 0.4403 & 0.3776 & 0.6807 & 0.9358 & 0.9701 & 0.8240 & 0.8353 & 0.7924 \\
% BERT4Rec & 0.2655 & 0.5747 & 0.7297 & 0.4265 & 0.4767 & 0.4119 & 0.6669 & 0.9230 & 0.9642 & 0.8094 & 0.8230 & 0.7787 \\ \midrule % (YELP 495.pkl)
% \modelname{}     & \textbf{0.2693} & \textbf{0.6065} & \textbf{0.7619} & \textbf{0.4448} & \textbf{0.4952} & \textbf{0.4247} & 0.7064 & 0.9316 & 0.9654 & 0.8326 & 0.8437 & 0.8056 \\
\bottomrule
\end{tabular}}
\label{tab:main_result}
\end{table*}

%\textbf{Evaluation Metrics.} 
\subsubsection{Evaluation Metrics} 
We adopt widely used top-$k$ Hit Ratio (HR@$k$), Normalized Discounted Cumulative Gain (NDCG@$k$), and Mean Reciprocal Rank (MRR) as metrics to evaluate the models. As HR@$1$ is equal to NDCG@$1$, we only report results on HR@$\{1,5,10\}$, NDCG@$\{5,10\}$ and MRR.
Following previous works~\cite{sun2019bert4rec,zhou2020s3}, we adopt leave-one-out strategy for evaluation. In particular, we use the last item of each sequence as the test data, and use the item before the last one as valid data. The rest items of the sequences are used as training data.
As the item set is quite large, it is very time-consuming to use all items as candidates to evaluate models. Therefore, we follow a common strategy~\cite{he2017neural,kang2018self} by randomly sampling $99$ negative items based on their popularity for each ground-truth item to speed up the experiments.

\subsubsection{Baseline Models}
To evaluate the effectiveness of our proposed model, we compare \modelname{} with following representative models.
\begin{itemize}
    \item[(1)] \textbf{GRU4Rec}~\cite{hidasi2016session} is a GRU-based model. It utilizes GRUs to model user behaviour sequences and adopts ranking based loss for session-based recommendation.
    \item[(2)]  \textbf{Caser}~\cite{tang2018personalized} employs convolutional neural network with both horizontal and vertical filters to capture sequential patterns from multiple levels, which allow it to model high order markov chains. % for sequential recommendation.
    \item[(3)] \textbf{SASRec}~\cite{kang2018self} utilizes self-attention mechanism for sequence modelling, which allows the model to capture long distance dependencies. It employs a left-to-right objective to optimize the model, and achieves promising results in next item recommendation task.
    \item[(4)] \textbf{BERT4Rec}~\cite{sun2019bert4rec} also adopts BERT to encode behaviour sequences. It proposes to use a cloze-style objective to generate representations with bidirectional context for sequential recommendation.
    \item[(5)] \textbf{$\text{\modelname{}}_{\text{w/o All}}$} is the pre-trained model with only the MIP task. The architecture of $\text{\modelname{}}_{\text{w/o All}}$ is same as \modelname{}, but it is pre-trained without user-aware tasks.
\end{itemize}

Notably, as previous pre-trained models mainly focus on utilizing item information or other recommendation tasks, these models cannot be applied in this paper. $\text{\modelname{}}_{\text{w/o All}}$ can serve as a strong pre-trained baseline.
% Notably, in the following ablation study, we present the performance of the pre-trained model with only the MIP task, which can serve as a strong baseline.

\subsubsection{Implementation Details}
For our proposed \modelname{}, we implement it by PyTorch and Transformers package~\cite{wolf2020transformers}. The hyper-parameters are selected by grid search on the valid dataset. We set the number of the transformers layers and the attention heads as $2$. The dropout rate is set as $0.5$. The dimension of the embeddings is set as $64$. For the YELP, the maximum length of behaviour sequences is set as $30$, and for the WeChat, the maximum length is set as $50$. Following \cite{zhou2020s3}, the mask proportion of item is set as $0.2$. In the pre-training stage, we set the weights for three loss (i.e., MIP, UAP, SRD) as $\lambda_1 = 1.0$, $\lambda_2 = 0.3$ and $\lambda_3 = 0.5$, respectively.  We employ Adam~\cite{kingma2015adam} as optimizer with the learning rate of $10^{-3}$. We set the batch size as $768$ for YELP, and $256$ for WeChat. We optimize the model with $1,500$ iteractions in each epoch. We pre-train our model for $75$ epochs and save checkpoint every $5$ epochs. Each checkpoint is further used to fine-tune for $40$ epochs, and the checkpoint with highest HR$@1$ scores on the valid datasets are used to evaluate on test datasets. In the fine-tuning stage, we set the learning rate as $10^{-4}$ and set the batch size as $256$.

For baseline models, we use the source code provided by the authors. For a fair comparison, we set the dimension of hidden vectors as $64$ for all baseline models. And for SASRec and BERT4Rec, which are self-attention based models, we set the number of model layers and attention heads as $2$. The remaining hyper-parameters are set following their suggestion in their papers. 

All models for the YELP dataset are trained on NVIDIA GeForce GTX 2080Ti GPUs, and models for the WeChat dataset are trained on NVIDIA Tesla P40 GPUs.

\subsection{Overall Performance Comparison}

The results of baseline models and \modelname{} are shown in Table~\ref{tab:main_result}. From the results we can observe that:

% 我们的模型在两个数据集上都要比之前的模型要好，说明我们的模型能够充分利用用户信息
Compared with the baseline models, \modelname{} can significantly outperform them by a large margin on both two datasets. The results show that our method can effectively incorporate various user information into pre-trained models, and generate expressive user representations based on their behaviour sequences. Moreover, both UPRec and BERT4Rec adopt BERT as the basic encoder and optimize the model with a cloze-style objective. And UPRec can achieve better performance for sequential recommendation, which further proves that constructing self-supervised signals from social networks and user attributes can help the model obtain general users' preferences and capture intricate sequence patterns.

%YELP数据规模较少的情况下，user-aware预训练任务相比之下提供了更多的额外信息，所以提升幅度比MLM都高

% self attention-based model 比前面两个model都要好，说明了self-attention机制拟合序列数据非常厉害
As for the baselines for sequential recommendation, we can observe that SASRec and BERT4Rec achieve better performance than Caser and GRU4Rec on the two datasets. Both SASRec and BERT4Rec employ the self-attention mechanism to capture information from behaviour sequences. This indicates that the self-attention mechanism is more suitable for sequence modelling than convolutional neural networks and recurrent neural networks.
% For sequential recommendation, attention based methods can significantly outperform Caser and GRU4Rec.

Moreover, the two attention-based models, SASRec and BERT4Rec, consist of the same model architecture, while the training objectives are different. SASRec adopts an autoregressive objective to train the model, which predicts the items unidirectionally. BERT4Rec adopts a cloze-style objective which can utilize bidirectional sequence information. BERT4Rec can consistently outperform the SASRec model, which indicates that it is important to generate representations bidirectionally for sequential recommendation.

\begin{table*}[h]
%\small
\centering
\caption{The performance on sequences with different length. The data is divided into three groups according to the length of behaviour sequences.}
\begin{tabular}{l|cc|cc|cc|cc}
\toprule
Dataset & \multicolumn{2}{c|}{small} & \multicolumn{2}{c|}{medium} & \multicolumn{2}{c|}{large} & \multicolumn{2}{c}{all} \\ \midrule
%\# Users & \multicolumn{2}{c|}{small} & \multicolumn{2}{c|}{medium} & \multicolumn{2}{c|}{large} & \multicolumn{2}{c}{all} \\ \midrule
Metrics     &  NDCG@10  &  MRR  &  NDCG@10  &  MRR  &  NDCG@10  &  MRR  &  NDCG@10  &  MRR \\ \midrule 
$\text{\modelname{}}_{\text{w/o All}}$ 
            &   36.47   & 29.66  & 37.86 &  30.96  & 38.11 & 30.95 & 37.14 & 30.25  \\
\modelname{}&   \textbf{39.19} ($\uparrow$ 2.71)  & \textbf{32.06} ($\uparrow$ 2.40)  & \textbf{40.28} ($\uparrow$ 2.42) &  \textbf{32.97} ($\uparrow$ 2.01) & \textbf{39.94} ($\uparrow$ 1.83) & \textbf{32.09} ($\uparrow$ 1.14) & \textbf{39.63} ($\uparrow$ 2.49) & \textbf{32.33} ($\uparrow$ 2.08)   \\ 
\bottomrule
\end{tabular}
\label{tab:split}
\end{table*}

To further evaluate how the two user-aware pre-training tasks improve the performance of recommendation, we show the results on behaviour sequences with different length. As shown in Table~\ref{tab:split}, we divide the data into three groups according to their sequence length. Here, sequences with length $< 8$ are divided into the small group, sequences with length $\geq 15$ are divided into the large group, and others are divided into the medium group. Due to the space limitation we report NDCG@$10$ and MRR for comparison. In order to investigate how the user-aware pre-training tasks benefit the performance, we compare the results of \modelname{} and $\text{\modelname{}}_{\text{w/o All}}$ in the experiments.
From the results, we can find that \modelname{} can achieve more improvements on the small group. It demonstrates that for the use with only a few interactions, the user attributes and social graphs can provide useful information, and help the model to capture their preferences more accurately. Besides, for sequences in the large group, even the behaviour sequences contain sufficient information, the extra user information can also further benefit the recommendation performance, which verifies the effectiveness of \modelname{} in integrating user information into the pre-trained model.

%\subsection{Further Analysis}
%Moreover, in order to further evaluate whether \modelname{} can work well, we conduct an ablation study, evaluate the performance on user profile prediction tasks and perform hyper-parameter sensitive analysis in this section.

\subsection{User Profile Prediction}
As our model aims to utilize the rich user information in pre-training, we argue that \modelname{} can achieve accurate user modelling. Therefore, we evaluate \modelname{} on the user profile prediction task. We adopt three tasks: (1) Compliment Prediction: it requires the model to prediction the number of compliments received by the user. (2) Average Star Regression: it requires the model to predict the average rating of reviews posted by the user. (3) Gender Prediction: it requires the model to predict the gender of the user. Notably, the first two tasks are also used in the pre-training stage, and gender prediction is a new challenging task, which is not adopted in pre-training. In these experiments, we employ the BERT as a baseline, which encodes the user behaviour sequences with BERT and utilizes the cross-entropy loss or Huber loss as objectives.

% compliment: 0.67653, star: 0.01961, gender: 0.61713
% compliment: 0.67521, star: 0.0375, gender: 0.60972 (without pre-training)
\begin{table}[h]
\small
\centering
\caption{The performance of user profile prediction task on the YELP dataset. We evaluate the performance on compliment prediction, average star regression, and gender prediction. We adopt accuracy as metric for compliment prediction task and gender prediction task. We adopt mean-square error as metric for the average star regression task.}
\begin{tabular}{c|ccc}
\toprule
Task    & Compliment &  Star & Gender \\ \midrule
BERT    &   0.6752   & 0.0375 & 0.6097 \\
UPRec   &   \textbf{0.6765}   & \textbf{0.0196} & \textbf{0.6171} \\ \bottomrule
\end{tabular}
\label{tab:user_pro}
\end{table}

The results are shown in Table~\ref{tab:user_pro}. We adopt accuracy as metric for compliment prediction and gender prediction. We adopt mean-square error as metric for average star regression. From the results, we can observe that \modelname{} can achieve performance improvements in all three tasks, especially in the average star regression task. Besides, though the gender prediction task is not used to pre-train our model, \modelname{} can also outperform the baseline model on this task. The improvements in first two tasks prove that our model can learn useful information from user attributes and social graphs, and thus benefit the recommender systems.
And the improvements in the gender prediction task demonstrate that our user-aware pre-training tasks can help the model to capture user attributes from their behaviours.
The results further verify that utilizing various user information in pre-training can significantly help the model to capture user preference effectively and accurately.

\subsection{Social Relation Detection}
\modelname{} adopts the social relation detection task to generate similar representations for social connected users. To verify whether \modelname{} can work well and generate social aware user representations, we evaluate our proposed model on social relation detection task. Specifically, as in pre-training stage, for each user in a social relation, we will sample $99$ negative candidate users and require the model to select the true friend according their behaviour sequences. We compare our model with two baselines: (1) Similarity (Sim): it assumes that friends tend to behave similarly and interact with the same items. Thus, it always predict the candidate user as the friend who has the most of same items with the query user. (2) BERT: it encodes the behaviour sequence with BERT, and is trained with the loss stated in Equation~\ref{eq:loss_srd}. 

\begin{table}[h]
\small
\centering
\caption{The performance of social relation detection task on the YELP dataset. We adopt the accuracy as the evaluation metric.}
\begin{tabular}{c|ccc}
\toprule
Model   &  Sim  &  BERT  &  UPRec \\ \midrule
Acc     & 12.87 &  69.43 &  \textbf{79.52} \\
\bottomrule
\end{tabular}
\label{tab:social_rel}
\end{table}

The results are shown in Table~\ref{tab:social_rel}. We adopt the accuracy as the evaluation metric. From the results, we can find that \modelname{} significantly outperforms the baseline models and are able to recommend friends for each user accurately, even the users are new for the model. It demonstrates that our model can effectively generate similar representations for friends, and thus benefit user preferences modelling. Besides, similarity strategy can also perform better than random prediction, which verify the hypothesis that friends tend to behave similarly and are supposed to have similar representations.

\subsection{Ablation Study}

\begin{table*}[t]
\centering
\caption{The results of ablation study.}
\resizebox{\textwidth}{!}{
\begin{tabular}{l|cccccc|cccccc}
\toprule
Dataset  & \multicolumn{6}{c|}{YELP}                    
         & \multicolumn{6}{c}{WeChat}   \\ \midrule
Metrics  & HR@1 & HR@5 & HR@10 & NDCG@5 & NDCG@10 & MRR
         & HR@1 & HR@5 & HR@10 & NDCG@5 & NDCG@10 & MRR \\ \midrule
BERT4Rec & 14.60 & 45.98 & 64.81 & 30.58 & 36.66 & 29.79 
         & 28.61 & 58.02 & 72.46 & 43.85 & 48.53 & 42.45 \\ \midrule % (source code)
w/o All  & 15.04 & 46.31 & 65.50 & 30.93 & 37.14 & 30.25 
         & 29.57 & 64.62 & 79.36 & 47.91 & 52.70 & 45.47  \\
w/o Rel  & 16.29 & 48.57 & 67.94 & 32.71 & 38.98 & 31.79
         & 29.82 & 64.84 & 79.45 & 48.15 & 52.90 & 45.69 \\
w/o Pro  & 16.46 & 48.29 & 67.67 & 32.62 & 38.89 & 31.77
         & 29.70 & \textbf{64.89} & \textbf{79.48} & 48.12 & 52.86 & 45.63 \\ \midrule
\modelname{}     & \textbf{16.96} & \textbf{49.04} & \textbf{68.81} & \textbf{33.24} & \textbf{39.63} & \textbf{32.31} 
        & \textbf{29.94} & 64.86 & 79.32 & \textbf{48.22} & \textbf{52.92} & \textbf{45.75} \\
\bottomrule
\end{tabular}}

\label{tab:ablation_study}
\end{table*}

To explore the contribution of two user-aware pre-training tasks, we conduct an ablation study and the results are shown in Table~\ref{tab:ablation_study}. Specifically, we show the scores with different pre-training tasks turned off. Here w/o All, w/o Rel, and w/o Pro refer to pre-training the model without user-aware tasks, social relation detection, and user attribute prediction, respectively. The results of the baseline with the best overall performance, BERT4Rec, are also provided for comparison.

From the results, we can observe that both two user-aware pre-training tasks contribute to the main model, as the performance decreases with any of the tasks missing. Note that the model without any user-aware pre-training tasks can also outperform BERT4Rec, which indicates that the two-stage pre-training and fine-tuning mechanism can improve the performance of the model. Besides, compared with the model pre-trained without user-aware tasks (w/o All), the models pre-trained with social relation detection (w/o Pro) or user attribute prediction (w/o Rel) can significantly achieve better results. It further proves that both two user-aware tasks can help the model to capture high order features and inject user information into the pre-trained models.
%Compared with the WeChat dataset, the YELP dataset contains sparser interactions, and thus the user information can provide more clues 

Notably, the results on the WeChat dataset show that the models with only one pre-training task (w/o Rel, w/o Pro) can achieve comparable performance on hit ratio. It demonstrates that the two pre-training tasks have a similar role for injecting the user information into representations to some extent. But pre-training with both two tasks can achieve more robust performance on various evaluation metrics. %The results on the YELP datasets show that 

The two datasets contain social graphs of different densities and different types of user attributes. And our proposed user-aware pre-training tasks can improve the performance significantly, which verify the effectiveness and robustness of our method.

%\IEEEpubidadjcol

%\subsection{Hyper-Parameter Sensitive Analysis}
\subsection{Performance w.r.t. the Number of Epochs}
\begin{figure}[h]
     \centering
     \begin{subfigure}
         \centering
         \includegraphics[width=0.23\textwidth]{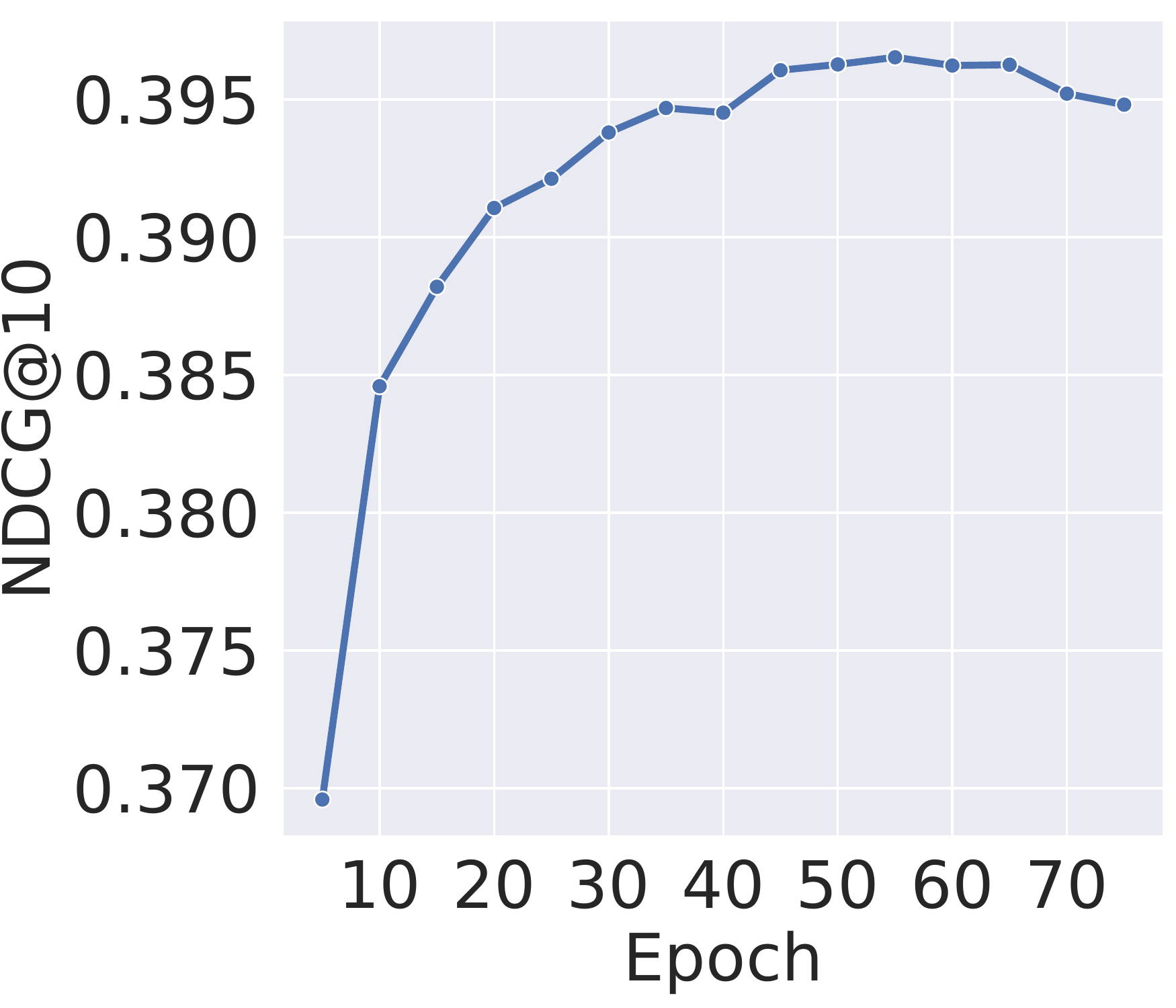}
         %\caption{$y=x$}
         \label{fig:y equals x}
     \end{subfigure}
     \begin{subfigure}
         \centering
         \includegraphics[width=0.23\textwidth]{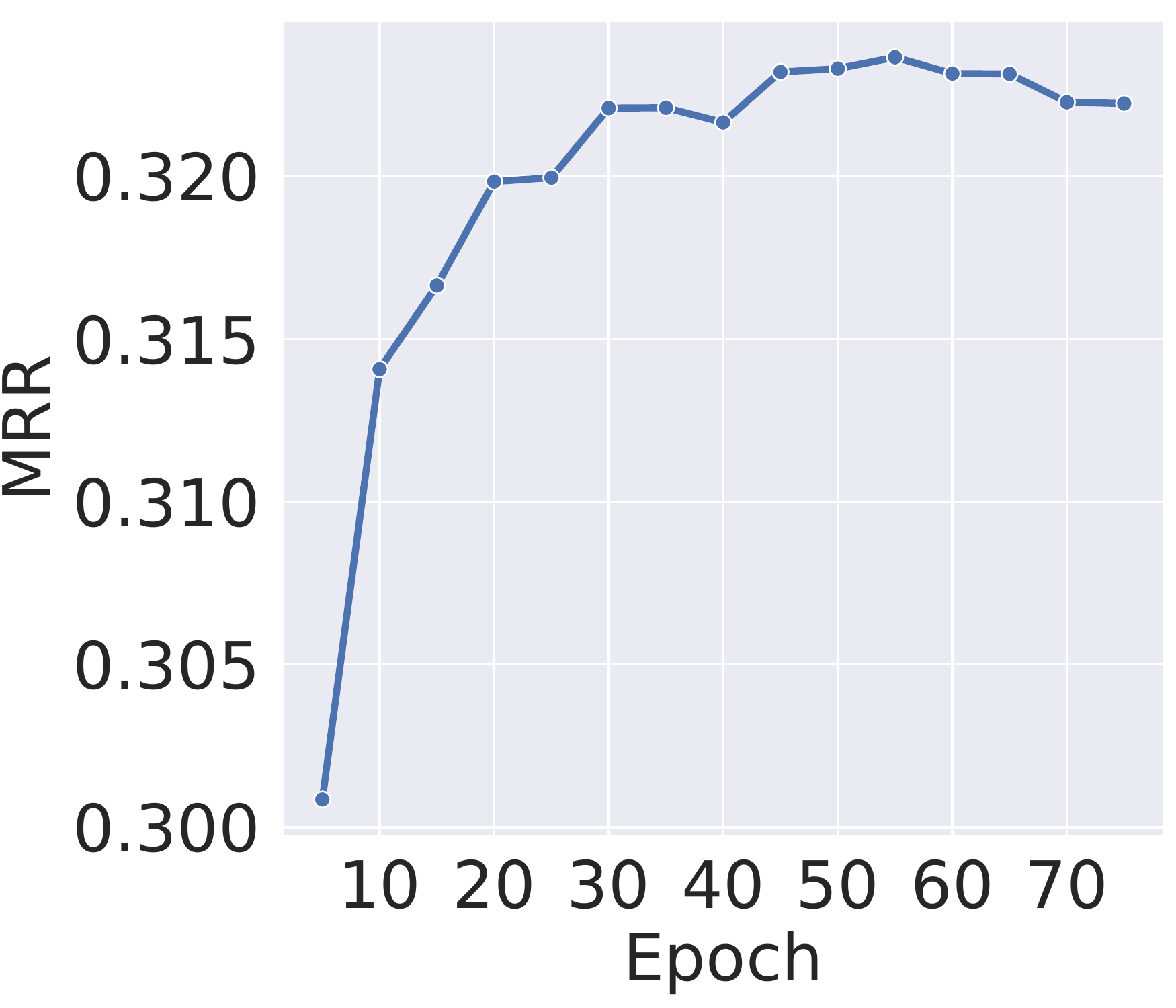}
         %\caption{$y=3sinx$}
         \label{fig:three sin x}
     \end{subfigure}
    \caption{The performance (NDCG@10 and MRR) with respect to the number of pre-training epochs on the YELP dataset.}
    \label{fig:performance_epoch}
\end{figure}
Our model consists of the pre-training stage and fine-tuning stage. In the pre-training stage, \modelname{} are trained to inject the user information into the user representations and item representations. The number of pre-training epochs can greatly influence model performance. Therefore, in order to study this issue, we pre-training the model with different number of epochs on the YELP dataset, and fine-tune them on the sequential recommendation task every $5$ epochs. 

Figure~\ref{fig:performance_epoch} represents the performance comparison with regard to the number of epochs on the YELP dataset. From the results, we can see that during the first $30$ epochs, the performance improves a lot, while after that the performance improves slightly. It proves that \modelname{} converges quickly and can effectively capture the features from the heterogeneous user information in the first few epochs. And thus the enriched representations are able to improve the performance of the sequential recommendation.

\subsection{Performance w.r.t. the Batch Size}
\begin{figure}[h]
     \centering
     \begin{subfigure}
         \centering
         \includegraphics[width=0.23\textwidth]{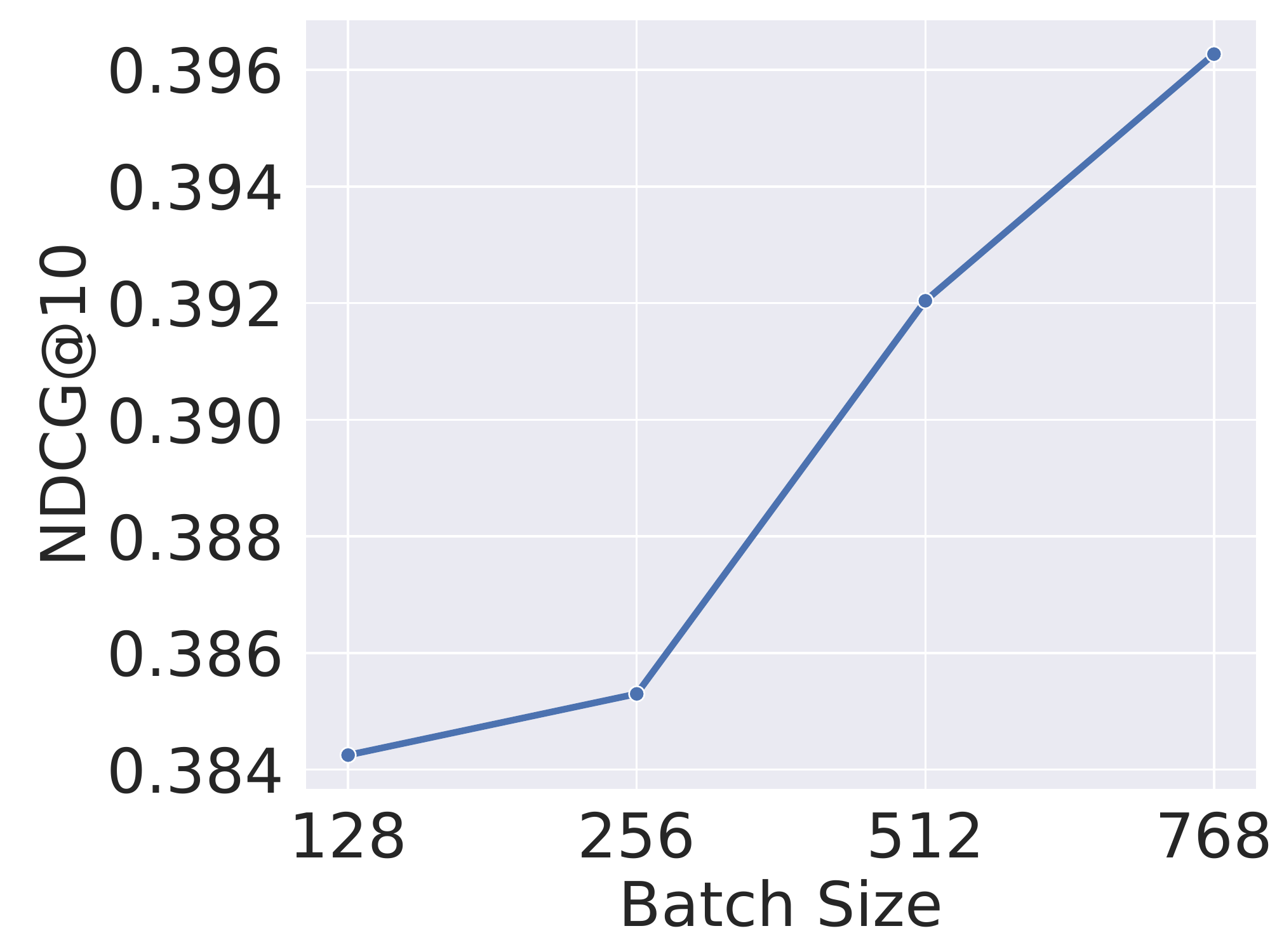}
         %\caption{$y=x$}
         \label{fig:y equals x}
     \end{subfigure}
     \begin{subfigure}
         \centering
         \includegraphics[width=0.23\textwidth]{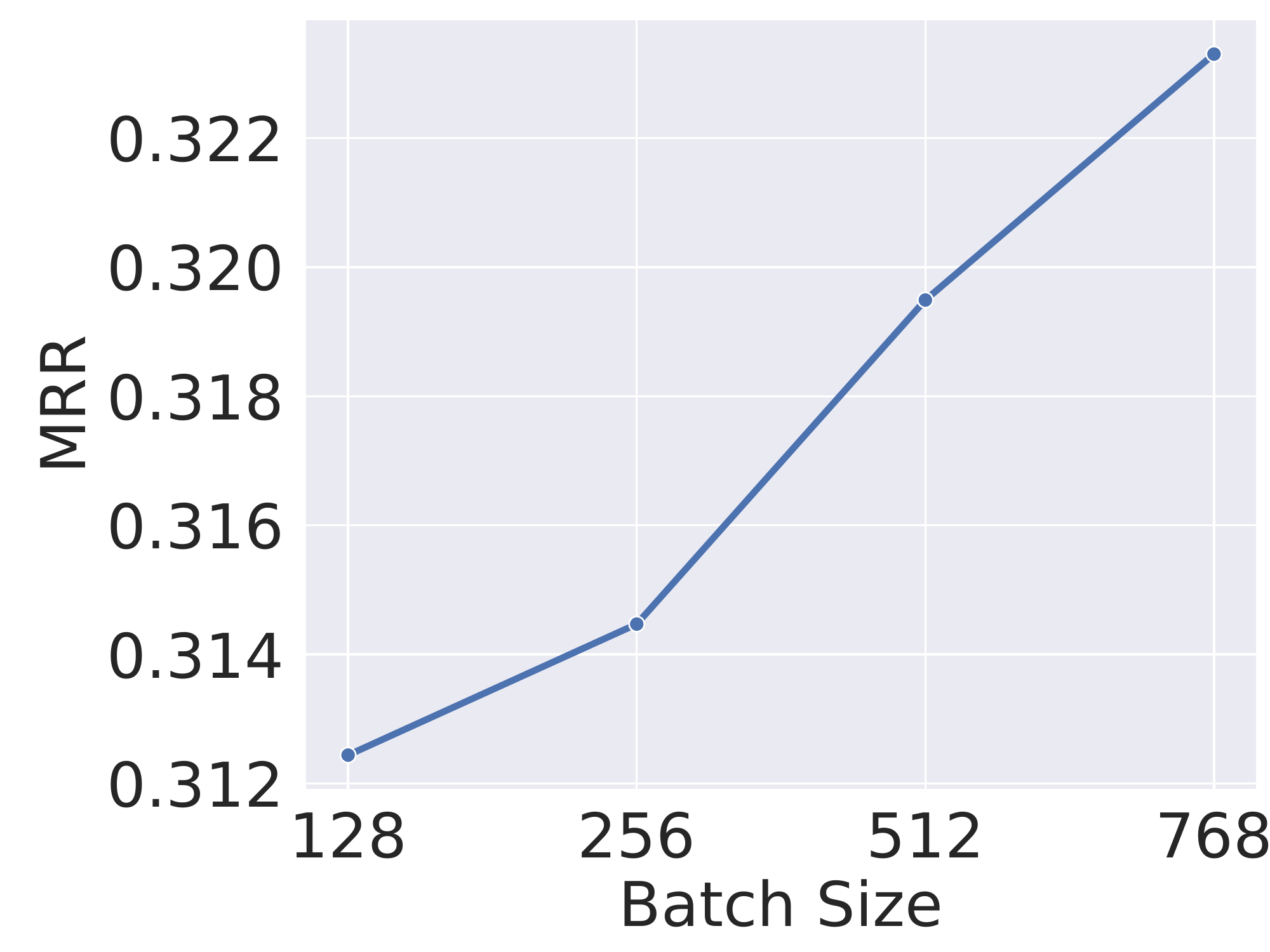}
         %\caption{$y=3sinx$}
         \label{fig:three sin x}
     \end{subfigure}
    \caption{The performance (NDCG@10 and MRR) with respect to the batch size on the YELP dataset.}
    \label{fig:performance_batchsize}
\end{figure}
It has been proven in the fields of NLP that hyper-parameter choices have a significant impact on the pre-trained models, and pre-training with bigger batch size can lead to better performance~\cite{liu2019roberta}. Therefore, we are wondering how the batch size of pre-training affects the performance. To investigate this, we pre-train our model with batch size as $\{128, 256, 512, 768\}$, and fine-tune them on the sequential recommendation task.

The results are shown in Figure~\ref{fig:performance_batchsize}. From the results, we can observe that the performance increases significantly with increasing the batch size. Large mini-batch can help the model to be optimized stably and efficiently. Besides, as we adopt the in-batch negative strategy for the social relation detection task, the larger batch size indicates the larger number of negative users. This can help the model generate expressive user representations, thus further improve the performance on downstream tasks. 
But we still have not reached the upper bound of the model's capacity. We can pre-train the model with a bigger batch size to achieve better performance, which we leave to future work.

\subsection{Performance w.r.t. the Hidden Size}
\begin{figure}[h]
     \centering
     \begin{subfigure}
         \centering
         \includegraphics[width=0.23\textwidth]{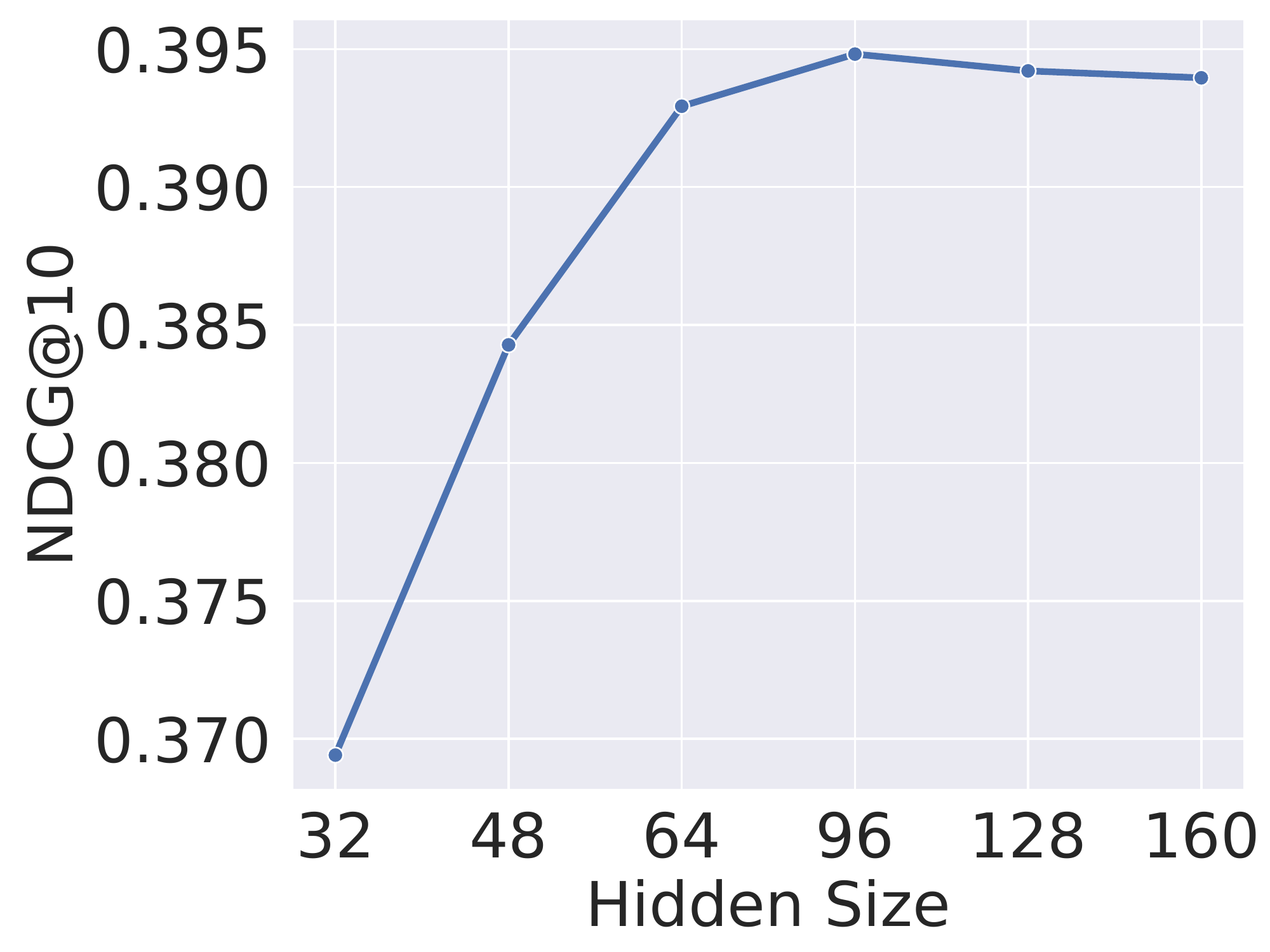}
         %\caption{$y=x$}
         \label{fig:y equals x}
     \end{subfigure}
     \begin{subfigure}
         \centering
         \includegraphics[width=0.23\textwidth]{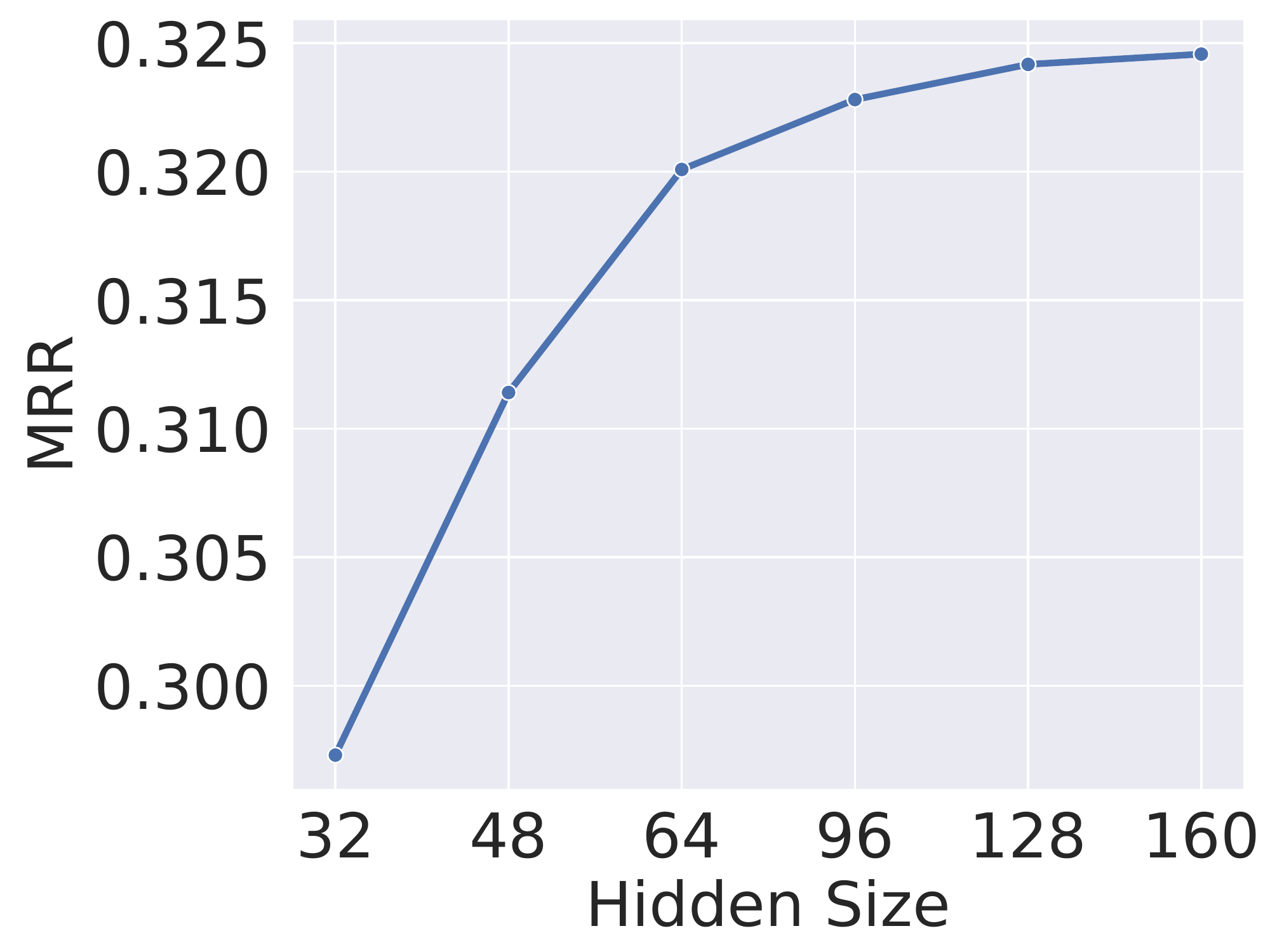}
         %\caption{$y=3sinx$}
         \label{fig:three sin x}
     \end{subfigure}
    \caption{The performance (NDCG@10 and MRR) with respect to the hidden size on the YELP dataset.}
    \label{fig:performance_hidden}
\end{figure}
Further, the number of parameters has a significant impact on the performance of pre-trained models. In the fields of pre-trained language models, it has been proven that a larger number of parameters can lead to better performance~\cite{devlin2019bert,liu2019roberta}. Therefore, in this part, we aim to study how the hidden size affects the recommendation performance and whether a larger model can lead to better performance in the recommendation. Figure~\ref{fig:performance_hidden} presents the NDCG@$10$ and MRR with the hidden size varying from $32$ to $160$ while keeping other hyper-parameters unchanged. From the results, we can observe that the performance benefits a lot when we increase the hidden size to $96$. After that, when we continue to increase the hidden size, the NDCG@$10$ score decreases. That is probably caused by overfitting. Therefore, we should choose the hidden size carefully for different datasets with various sparsity and scale.

To summarize, from the results of hyper-parameter sensitive analysis, we can conclude that the pre-training mechanism can enable the model to recommend items accurately, even with the limited behaviour data. Besides, the results suggest that we can train large-scale pre-trained models with large batch size and large number of parameters to further improve the performance.

\section{Conclusion}
In this paper, we propose to incorporate user information into pre-trained models for recommender systems. In our model, we propose two novel user-aware tasks, including user attribute prediction and social relation detection, which are designed to utilize user attributes and social graphs. Then we evaluate our proposed \modelname{} on the sequential recommendation task and user profile prediction tasks. The experimental results demonstrate that our model can generate expressive user representations from their behaviour sequences, and outperform other competitive baseline models. Besides, we conduct an ablation study and hyper-parameter sensitive analysis, which suggest that pre-training with user-aware tasks can improve the performance, and we can train large models with large batch size to further promote the progress.

In the future, we will explore to how to design powerful pre-training tasks to further utilize more user information, including their posted reviews and other behaviours. 
It is also worthy to explore how our model perform in other complex recommendation tasks, such as next basket recommendation and click through rate prediction.

% or


% \appendices

% \section{}

\ifCLASSOPTIONcompsoc
  \section*{Acknowledgments}
\else
  \section*{Acknowledgment}
\fi

This work is supported by the National Key Research and Development Program of China (No. 2020AAA0106501) and the National Natural Science Foundation of China (NSFC No. 61772302).
Yao is also supported by 2020 Tencent Rhino-Bird Elite Training Program.

\ifCLASSOPTIONcaptionsoff
  \newpage
\fi



%\bibliographystyle{IEEEtran}
\bibliographystyle{IEEEtran}
\bibliography{bare}

\end{document}